\documentclass[12pt,preprint]{aastex}

\slugcomment{Not to appear in Nonlearned J., 45.}

\shorttitle{Accretion Properties of Nearby Hard X-ray AGNs}
\shortauthors{Wang \& Wei}

\begin{document}


\title{Accretion Properties of 
   A Sample of Hard X-ray ($<60$keV) Selected Seyfert 1 Galaxies}


\author{J. Wang\altaffilmark{1} Y. F. Mao \altaffilmark{1} and J. Y. Wei\altaffilmark{1} }
\affil{National Astronomical Observatories, Chinese Academy of Science, 20A Datun Road,
Chaoyang District, Beijing, China}
\email{wj@bao.ac.cn}



\begin{abstract}
We examine the accretion properties in 
a sample of 42 hard (3-60keV) X-ray selected nearby broad-line AGNs. The energy range 
in the sample is harder than that usually used in the similar previous studies.
These AGNs are mainly complied from the RXTE
All Sky Survey (XSS), and complemented by the released \it INTEGRAL \rm
AGN catalog. The black hole masses, bolometric luminosities of AGN, and
Eddington ratios are derived from 
their optical spectra in terms of the broad H$\beta$ emission line. The tight correlation 
between the hard X-ray (3-20keV) and bolometric/line luminosity is
well identified in our sample. Also identified 
is a strong inverse Baldwin relationship of the H$\beta$ emission line.
In addition, all these hard X-ray AGNs are biased toward luminous objects with 
high Eddington ratio (mostly between 0.01 to 0.1) and low column density ($<10^{22}\ \mathrm{cm^{-2}}$),
which is most likely due to the selection effect of the surveys.
The hard X-ray luminosity is consequently found to be strongly 
correlated with the black hole mass. 
We believe the sample completeness 
will be improved in the next few years by the ongoing \it Swift \rm and \it INTEGRAL \rm missions, 
and by the next advanced missions, such as NuSTAR, Simbol-X, and NeXT.  
Finally, the correlation between 
RFe (=optical \ion{Fe}{2}/H$\beta$) and disk temperature as assessed by 
$T\propto (L/L_{\mathrm{Edd}})M_{\mathrm{BH}}^{-1}$ leads us 
to suggest that the strength of the \ion{Fe}{2} emission is mainly determined by the shape of the ionizing spectrum.  
\end{abstract}





\section{Introduction}

It is now generally believed that the power of Active Galactic Nuclei (AGNs)
is extracted through the accretion of gas onto a central supermassive 
black hole (SMBH). Such energy mechanism means that AGNs are characterized by 
their strong hard X-ray emission ($h\nu >2$keV), which is widely used as direct evidence 
suggesting the existence of a nuclear accretion activity 
($L_\mathrm{2-8keV}>10^{42}\ \mathrm{ergs\ s^{-1}}$, e.g., Silverman et al. 2005; Brusa 
et al. 2007; Hasinger et al. 2005).
The commonly accepted model is that the hard X-ray emission from AGNs is 
primarily produced by the inverse Compton scattering of the UV/soft X-ray photons emitted from 
the accretion disk (e.g, Zdziarski et al. 1995, 2000; Haardt \& Maraschi 1991; Kawaguchi 
et al. 2001). The absorption-corrected X-ray photon spectra within the energy band 2-10keV 
could be best described as a cut-off powerlaw with index $\Gamma\sim1.9$ (e.g., 
Zdziarski et al. 1995; Reeves \& Turner 2000; Piconcelli et al. 2005; Dadina 2008; 
Panessa et al. 2008). The synthesis spectra become flat beyond 10keV because 
of the Compton reflection caused by the ionized surface of the accretion disk 
(e.g., George \& Fabian 1991).

Because the hard X-ray emission from central engine can penetrate the obscuration material 
much more easily than lower energy emission, it possesses particular importance in
testing the traditional unified model (Antonucci 1993) in Seyfert 2 galaxies (e.g., 
Moran et al. 2002; Cardamone et al. 2007). 
The studies of the \it Chandra \rm and \it XMM-Newton \rm observatories showed that the 
cosmic X-ray background (CXRB) at 2-30keV
might be contributed by many unknown obscured AGNs which are predicted by the
CXRB models (e.g., Comastri et al. 1995; De Luca \& Molendi 2004; Worsley et al. 2005;
Gilli et al. 2007; Severgnini et al. 2003; Levenson et al. 2006).
Taking into account of the issue of co-evolution of AGN
and bulge of its host galaxy (e.g., Heckman et al. 2004; Kauffmann et al. 2003;
Wang et al. 2006; Wang \& Wei 2008 and references therein), 
the hard X-ray emission from AGNs is also
an important tool in detecting and separating AGN's contribution from circumnuclear 
star formation activity. The X-ray luminosities of 
known most X-ray luminous starforming and elliptical galaxies are not higher than
$L_{\mathrm{X}}=10^{42}\ \mathrm{erg\ s^{-1}}$ (Zezas et al. 2003; Lira et al. 2002a, 2002b;
O'Sullivan et al. 2001).

Heckman et al. (2005) identified a very tight correlation between the hard X-ray (3-20keV)  
and [\ion{O}{3}] luminosities in a sample of hard X-ray selected AGNs 
when they performed a comparison between the hard X-ray
selected and [\ion{O}{3}] emission-line selected AGNs. In 2-10keV bandpass, 
similar correlations were identified in the Palomar optically selected AGNs 
by Panessa et al. (2006).
On the contrary, a very weak $L_{\mathrm{[OIII]}}$-$L_{\mathrm{X}}$ correlation was
identified in the AGNs selected by their bright [\ion{O}{3}] emission lines 
(Heckman et al. 2005). The result suggests that many AGNs might be missed in the 
hard X-ray survey. Moreover, Netzer et al. (2006) claimed that the 
$L_{\mathrm{[OIII]}}$/$L_{\mathrm{X}}$ ratio depends on the X-ray luminosity.

The questions are therefore naturally raised: why is the $L_{\mathrm{[OIII]}}$-$L_{\mathrm{X}}$
correlation broken in some kind of AGNs? which parameters (or what are the physical reasons that)
determine the correlation? Are the hard X-ray selected AGNs particular in some parameters?
Both black hole mass ($M_{\mathrm{BH}}$) and Eddington ratio ($L/L_{\mathrm{Edd}}$) are
two key parameters determining the observed properties of AGNs. In addition, with the 
development of the technology in hard X-ray detection, a major advance in studying AGN hard X-ray emission 
will be achieved in the next 
a few years due to the launch of new missions with enhanced hard X-ray capability (in sensitivity 
and imaging), such as NuSTAR, Simbol-X, and NeXT (e.g., Takahashi et al. 2008; Ferrando et al. 2003). 
The study on the existent surveys certainly prepares the ground for the future surveys.

In this paper, we examine the optical spectral properties of a sample of 42 hard X-ray 
selected broad-line AGNs, which allows us to investigate the properties of 
$M_{\mathrm{BH}}$ and $L/L_{\mathrm{Edd}}$ in these hard X-ray AGNs. 
The sample is mainly compiled from the AGN catalog of the RXTE 3-20keV All Sky Survey (XSS, 
Sazonov \& Revnivtsev 2004), and complemented by the AGN catalog 
released by the \it INTEGRAL \rm all-sky hard X-ray survey (Bassani et al. 2006a and references therein). 
Note that the energy range used in this work is extended to the Compton reflection region, 
and is harder than the range (i.e., 0.5-10 keV) usually used in the similar previous studies.  
The paper is organized as follows. Section 2 and 3 shows the sample selection and spectroscopic 
observations, respectively.
The data reduction is described in the next section. Section 5 presents the 
results and discussions. The cosmology with $h_0=0.7$, $\Omega_M=0.3$, and $\Omega_\Lambda=0.7$ 
(Bennett et al. 2003) is adopted in our calculations throughout the paper.

\section{The Sample}





Our sample is mainly compiled from the RXTE 3-20keV All-Sky Survey 
(XSS, Revnivtsev et al. 2004). Sazonov \& Revnivtsev (2004, hereafter SR04) identified 
95 nearby AGNs with 
$z_{\mathrm{median}}\sim0.035$ detected in XSS. The survey covers about 90\% of the sky
at $|b|>10\symbol{23}$. The sensitivity of the survey in 
the energy band 3-20 keV is better than $2.5\times10^{-11}\mathrm{ergs\ s^{-1}\ cm^{-2}}$.
SR04 therefore provides a nearly complete sample for bright hard X-ray selected 
AGNs, although the sample is still biased against sources with absorption column 
$N_{\mathrm{H}}\geq 10^{23}\mathrm{cm^{-2}}$. The IBIS telescope (Ubertini et al. 2003)
onboard the  \it INTEGRAL \rm observatory (Winkler et al. 2003) provides a better 
capability for heavily obscured objects because of its good sensitivity beyond 20 keV. 
After about three years \it INTEGRAL \rm observation, their all-sky survey project
allows Bassani et al. (2006a) to 
identify 62 AGNs in the energy bandpass 20-100keV above a flux of 
$1.5\times10^{-11}\mathrm{ergs\ s^{-1}\ cm^{-2}}$ by analyzing 11,300 
\it INTEGRAL \rm points (see also in Bassani et al. 2006b; Bird et al. 2006; 
Beckmann et al. 2006).

Both catalogs are then combined to enlarge our sample content, and to attempt to 
alleviate the bias against the obscuration.   
We further limit the objects to the broad-line AGNs 
according to the identification provided by 
previous optical spectroscopy. The BAT instrument (Barthelmy et al. 2005) onboard the 
\it Swift \rm satellite (Gehrels et al. 2004) provides an all-sky hard X-ray survey with a sensitivity 
down to a few $\times 10^{-11}\mathrm{ergs\ s^{-1}\ cm^{-2}}$ in higher energy band 14-195 keV. 
Markwardt et al. (2005) identified 44 previous known AGNs in the \it Swift \rm survey. 
However, considering the spectroscopic observation condition (see below), only three objects 
are not listed in our sample. It is worthy noting that the ongoing \it Swift\rm/BAT and 
\it INTEGRAL \rm survey could provide a larger sample of $\sim100$ hard X-ray selected AGNs at 
the end of our program (i.e., Krivonos et al. 2006; Sazonov et al. 2007; Tueller et al. 2008). 
We will perform a subsequent spectroscopic study on these AGNs in the next work. 

Figure 1 shows the column density ($\mathrm{N_H}$) distribution for our sample. 
$\mathrm{N_H}$ is complied from SR04, and from Sazonov et al. (2007) for the objects 
detected by \it INTEGRAL \rm only. Briefly, these authors determined $\mathrm{N_H}$ by 
fitting the available spectra from different X-ray instruments by an absorbed powerlaw.
Values of $\mathrm{N_H}$ less than $10^{22}\ \mathrm{cm^{-2}}$ were ignored in the 
fitting of SR04 and Sazonov et al. (2007). As 
shown in the figure, the current sample is still strongly biased against AGNs with large 
column density.

\section{Spectroscopic Observations}

Both because of the constraint of the observatory site and because of the instrumental capability, 
the spectroscopic observations are only carried out for the bright ($m_v<16.5$mag) objects
located in the northern sky with declination $\delta>-20\symbol{23}$. In total, 
there are 53 objects fulfilling the selection criterion. Six of these objects 
are not observed because of the poor weather conditions. Among the remaining objects, 
five objects (i.e., XSSJ\,02151-0033, XSS\,J11570+5514, XSS\,J22363-1230, NGC\,788, IGR\,J21247+5058)
are excluded in subsequent spectral modeling because of the weakness or absence of broad H$\alpha$ and/or
H$\beta$ emission lines\footnote{Our spectroscopic observations show that three objects, i.e., 
XSS\,J02151-0033, XSS\,J11570+5514, and XSS\,J22363-1230, could be further classified 
as Seyfert 1.9-like galaxies without broad components of Balmer emission lines except a broad 
H$\alpha$ emission. In addition, the continuum is dominated by the contribution from starlight of  
host galaxy rather than AGN in XSS\,J02151-0033 and XSS\,J11570+5514, which precludes the measurements 
of the \ion{Fe}{2} complex. We
include NGC\,788 in our observation program because multi-classification is shown in the
NED (see also in Beckmann et al. 2006). Our spectrum indicates a Seyfert 2 nucleus in the object.
In fact, the object is classified as a Seyfert 2 galaxy with 
polarized broad emission lines in the catalogue of quasars and active nuclei (Veron-Cetty \& Veron 2006). 
Our spectrum taken in Dec 20, 2006 shows that the spectrum of IGR\,J21247+5058 is 
dominated by a typical late type star, which is consistent with Masetti et al. (2004). 
By identifying the broad emission bump around 6700\AA\ as H$\alpha$, these author argued that the background AGN 
is aligned with a F- or early G-type star by chance.}

The total 42 high (or intermediate high) quality optical spectra were obtained by using the 
National Astronomical Observatories, Chinese Academy of Science (NAOC), 2.16m 
telescope in Xinglong observatory during several observing runs carried out 
from November 2005 to Mar 2008. The spectra were taken by the OMR spectrograph 
equipped with a back-illuminated SPEC10 1340$\times$400 CCD as detector. A grating of 
300$\mathrm{g\ mm^{-1}}$ and a slit of 2.0\symbol{125} oriented in south-north
direction were used in our observations. This setup results in a final spectral 
resolution $\sim 9\AA$ as measured from both comparison spectra and night sky 
emission lines. The blazed wavelength was fixed to be 6000\AA, which provides a 
wavelength coverage of 3800-8300\AA\ in observer frame. This attempt covers both 
H$\alpha$ and H$\beta$ region in most spectra because of their 
small redshifts. Each object was observed successively twice. 
The two exposures were combined prior to extraction to enhance the S/N ratio and
eliminate the contamination of cosmic-rays easily. The exposure time for each 
frame is generally between 900 and 3600 seconds depending on the brightness
of the object. The wavelength calibration associated with each object was
carried out by the helium-neon-argon comparison arcs taken between the two successive frames.
The arcs were obtained at the position being nearly identical to that of the specified object.  
Two or three Kitt Peak National Observatory (KPNO) standard stars (Massey et al. 1988)
were observed per night for both flux calibration and removal of the 
atmospheric absorption features. All the objects were observed as close to meridian 
as possible. Table 1 shows the log of observations of the 42 objects listed in our 
sample.

\section{Data Reduction and Emission Line Measurements}

The standard procedures using the IRAF package \footnote{IRAF is distributed by the 
National Optical Astronomical Observatories, which is operated by the Association 
of Universities for Research in Astronomy, Inc., under cooperative agreement with the 
National Science Foundation.} are adopted by us to reduce the 
unprocessed frames. The CCD reductions include bias subtraction, flat-field correction,
and cosmic-rays removal before the extraction of signal. One-dimensional sky-subtracted 
spectra are then wavelength and flux calibrated. The uncertainties of the wavelength and flux 
calibrations are no more than 1\AA\ and 20\%, respectively. The Galactic extinctions 
are corrected by the color excess, the parameter $E(B-V)$ taken from the NASA/IAPC Extragalactic
Database (NED), assuming an $R_V=3.1$ extinction law of Milk Way (Cardelli et al. 1989). 
The spectra are then transformed to the rest frame, along with a $k$-correction,  
according to the narrow peak of H$\beta$.       
 
\subsection{\ion{Fe}{2} subtractions}

In order to determine the strength of the optical \ion{Fe}{2} complex and to
reliably profile the emission lines in H$\beta$ region, the \ion{Fe}{2} complex should 
be modeled and then removed from each spectrum at first. We 
simply adopt the empirical template of the \ion{Fe}{2} complex introduced in 
Boroson \& Green (1992, hereafter BG92). For each object,  
the template is first broadened to the FWHM of the H$\beta$ broad component 
by convolving with a Gaussian profile (BG92).
The broadened template and a broken powerlaw are combined to model the continuum by a $\chi^2$
minimization for individual object. The minimization is performed over the 
rest frame wavelength range from 4300 to 7000\AA, except for the regions around 
the strong emission lines (i.e., H$\alpha$+[\ion{N}{2}], [\ion{S}{2}], H$\beta$, 
H$\gamma$, [\ion{O}{3}], \ion{He}{2}). The fittings
are illustrated in the left column of Figure 2 for three typical cases.
The flux of the \ion{Fe}{2} complex is integrated between the rest 
wavelength 4434 and 4686\AA. 
The \ion{Fe}{2} complex is not measured in XSS\,J11067+7234 because its continuum  
is dominated by the starlight from the host galaxy. 

\subsection{Emission lines measurements}

After removing the \ion{Fe}{2} blends and continuum, the isolated AGN emission lines
at H$\beta$ region are modeled by the SPECFIT task (Kriss 1994) in IRAF package. 
In each object, each of the [\ion{O}{3}]$\lambda\lambda$4959,5007 doublet is modeled 
by a single Gaussian component. The flux ratio of the doublet is fixed to be 3 
(e.g., Dimitrijevic et al. 2007). In principle, each H$\beta$ profile is modeled by a set 
of two Gaussian components: a narrow and a broad Gaussian component (denoted as 
H$\beta_\mathrm{n}$ and H$\beta_\mathrm{b}$, respectively). The profile modelings are 
illustrated for the three cases in the right column in Figure 2. The deblendings of the 
H$\beta$ emission line are, however, difficult in 15 objects because no obvious transition
between the narrow and broad components is observed. In these 15 cases, the reported 
H$\beta_\mathrm{b}$ measurements include both components, and the 
contribution of H$\beta_\mathrm{n}$ is usually expected to be less than 3\% for typical broad-line 
AGNs as suggested by BG92. 

The described modeling works very well in most objects, except in three ones. 
In XSS\,J22539-1735, a very broad H$\beta$ component ($\mathrm{FWHM\sim10,000\ km\ s^{-1}}$) 
is required to properly reproduce the observed profile. Such component has been 
reported in H$\beta$ and highly ionized emission lines in previous studies 
(e.g., Sulentic et al. 2000b; Wang et al. 2005; Veron-Cetty et al. 2007; 
Marziani et al. 2003; Mullaney \& Ward 2008; Hu et al. 2008). A double 
peaked and a boxy H$\beta$ profile is observed in XSS\,J18408+7947 and XSS\,J23073+0447,
respectively. In all of the three cases, each H$\beta$ line is modeled by three 
Gaussian components, i.e., a narrow and two broad components. The modeled narrow 
component is then subtracted from the observed spectrum to derive a residual profile 
which is then used to measure FWHM and integrated flux.  
 
\section{Results and Discussion}

Table 2 lists the following items measured from the spectra. Column (2) and (3)
lists the equivalent widths (EWs) of H$\beta_{\mathrm{n}}$ and H$\beta_{\mathrm{b}}$,
respectively. All the EWs refer to the continuum level determined from the continuum fitting 
at the position of wavelength $\lambda$4861. The line ratio RFe (=\ion{Fe}{2}/H$\beta_b$) 
is shown in column (4). Column (5) lists the luminosity of the
H$\beta$ broad component, and Column (6) the FWHM of the broad H$\beta$ in units of $\mathrm{km\ s^{-1}}$. 
All quoted widths are not corrected for the intrinsic resolution, since the widths are 
much broader than the resolution. Column (7) shows the luminosity of the [\ion{O}{3}] emission line.
Both [\ion{O}{3}] and H$\beta_b$ luminosities are calculated from the cosmology-corrected luminosity distance
\footnote{Because the objects listed in our sample are generally very nearby, the radial velocity is
corrected by a redshift of 0.017877 to the reference frame defined by the 3K Microwave Background Radiation.
}
given by NED. The next two columns show the absorption-corrected hard X-ray luminosities 
in the bandpass 3-20 keV (XSS) and in 17-60 keV (\it INTEGRAL\rm).

Six out of the 42 AGNs are detected by \it INTEGRAL \rm only. In order to estimate their X-ray luminosities
in the bandpass 3-20 keV from the luminosities in 17-60 keV, a transformation is derived in terms of the 
18 objects detected by both surveys. The left panel of Figure 3 shows the relationship between the two 
sets of luminosity. A unweighted fitting yields a relationship 
$\log L_{\mathrm{17-60 keV}}=1.03\log L_{\mathrm{3-20keV}}-1.39$ with a standard deviation of 0.25.
After transform the luminosities in the bandpass 17-60 keV to 3-20 keV for the six objects,   
the strong correlation between the [\ion{O}{3}] and 3-20 keV luminosity is shown in the right panel of Figure 3 
for the total 42 objects. The correlation is highly consistent with that derived in Heckman et al. (2005), 
which firmly demonstrates the accuracy of our observations and calibrations.  

\subsection{Hard X-ray vs. bolometric luminosity}

The detection of hard X-ray emission from a nucleus is regarded as strong evidence of accretion activity 
occurring around the central SMBH.
The correlations between X-ray luminosity and luminosities of optical emission lines (e.g., H$\alpha$, 
H$\beta$, [\ion{O}{3}]) have been extensively established in the previous studies 
(e.g., Elvis et al. 1984; Ward et al. 1988; 
Mulchaey et al. 1994; Heckman et al. 2005; Panessa et al. 2006). However, the correlations seem to 
depend on sample selection (Heckman et al. 2005) and on X-ray luminosity itself (Netzer et al. 2006).
In this paper, we estimate the  
luminosity of AGN at rest wavelength 5100\AA\ from the H$\beta_b$ component according to the calibration 
given by Greene \& Ho (2005):
\begin{equation}
L_{5100}=7.31\times10^{43}\bigg(\frac{L_{\mathrm{H\beta}}}{10^{42}\ \mathrm{erg\ s^{-1}}}\bigg)^{0.883}\ \mathrm{ergs\ s^{-1}} 
\end{equation}
The bolometric luminosity is then derived by adopting the usually used bolometric correction 
$L_{\mathrm{bol}}\approx 9\lambda L_{\lambda}(5100\AA)$ (e.g., Kaspi et al. 2000). The significant
correlation between $L_{\mathrm{bol}}$ and $L_{\mathrm{3-20keV}}$ is shown in the left panel in Figure 4. 
A unweighted least square fitting gives a relationship:
\begin{equation} 
\log L_X = (0.91\pm0.06)\log L_{\mathrm{bol}}+(3.04\pm2.78)
\end{equation}
or $L_{\mathrm{bol}}/L_X\sim 10$ for simplification. The correlation strongly 
indicates a close link between the hard X-ray emission and ionizing radiation emitted from the accretion disk. 
Note that the bolometric luminosity spans four orders of magnitude, down to $L_{\mathrm{bol}}\sim10^{43}\ 
\mathrm{ergs\ s^{-1}}$, which is close to the formal definition of low luminous AGNs 
(i.e., $L_{\mathrm{bol}}<10^{43}\ \mathrm{ergs\ s^{-1}}$, Ho 2003). These faint galaxies generally 
radiate at low state with Eddington ratio marginally exceeds 0.01.

\subsection{Black Hole Mass and Eddington Ratio}

It is now generally believed that black hole mass ($M_{\mathrm{BH}}$) and specific accretion rate 
($L/L_{\mathrm{Edd}}$) are two basic parameters determining the properties of AGNs. The 
role of $L/L_{\mathrm{Edd}}$ in driving the Eigenvector I (EI) space 
\footnote{ The EI space is one of the key properties of AGN
phenomena. It was first established by BG92 who analyzed the optical spectra of 87 bright PG quasars. 
In addition to the anti-correlation between the intensity of \ion{Fe}{2} and [\ion{O}{3}],
the EI space has been subsequently extended to ultraviolet and soft X-ray bands (i.e., FWHM(\ion{C}{4}) and
$\Gamma_\mathrm{s}$, e.g., Wang et al. 1996; Xu et al. 2003; Grupe 2004; Sulentic et al. 2007;
see Sulentic et al. 2000a for a review). 
}
has been extensively investigated in 
numerous previous studies (e.g., Boroson 2002; Sulentic et al. 2006; Xu et al. 2003; Grupe 2004; 
Zamanov \& Marziani 2002; Marziani et al. 2001), because of the great progress in the calibration of 
the $R_{\mathrm{BLR}}-L$
relationship (e.g, Kaspi et al. 2000, 2005, 2007; McLure \& Jarvis 2004; Vestergaard \& Peterson 2006; 
Peterson et al. 2004; Bentz et al. 2006) due to the recent great advance in 
the reverberation mapping (e.g., Kapsi et al. 2000; Peterson \& Bentz 2006). 
We refer the reads to McGill et al. (2008, and references 
therein) for a summary of the existing formula used to calculate $M_{\mathrm{BH}}$ basing upon 
``single-epoch'' observation. In this work, the black hole mass is estimated 
by the width and luminosity of H$\beta_b$ in individual object, according to the 
scaling law obtained by Greene \& Ho (2005).  
\begin{equation}
M_{\mathrm{BH}}=3.6\times10^6\bigg(\frac{L_{\mathrm{H\beta}}}{10^{42}\ \mathrm{erg\ s^{-1}}}\bigg)^{0.56}
\bigg(\frac{\mathrm{FWHM_{H\beta}}}{1000\ \mathrm{km\ s^{-1}}}\bigg)^2\ \mathrm{M_{\odot}}
\end{equation}
For each object, the estimated $M_{\mathrm{BH}}$ and $L/L_{\mathrm{Edd}}$ is listed in Column (10) and (11) 
in Table 2, respectively. The bolometric luminosity $L_{\mathrm{bol}}$ is estimated from the
H$\beta_b$ component as described above.   

The calculations of both $L_{\mathrm{bol}}$ and $L/L_{\mathrm{Edd}}$ allow us to 
find that the current sample is strongly biased against sub-luminous AGNs usually with 
low $L/L_{\mathrm{Edd}}$. In fact, the left panel of Figure 4 shows the lack of AGNs with 
$L_{\mathrm{bol}}<10^{43}\ \mathrm{ergs\ s^{-1}}$.  
Figure 5 displays the distributions of the $L/L_{\mathrm{Edd}}$ (\it left panel\rm) 
and $M_{\mathrm{BH}}$ (\it middle panel\rm) for the total 42 AGNs. 
As shown in the middle panel, the $M_{\mathrm{BH}}$ is sampled within a wide range ($\sim3$ dex) 
from $10^{6}M_\odot$ to $10^{9}M_\odot$ with a peak at $10^{7-8}M_\odot$. In contrast, the left panel 
shows that $L/L_{\mathrm{Edd}}$ distributes in a relatively narrow range as compared with the previous studies. 
The total range of our $L/L_{\mathrm{Edd}}$ spans from 0.01 to 1. 
In particular, about $\sim60$\% objects listed in the sample have $L/L_{\mathrm{Edd}}$ between 0.01 and 0.1.
However, the $L/L_{\mathrm{Edd}}$ of bright local AGNs usually spans at least three orders of 
magnitude from 1 to 0.001 (e.g., Woo \& Urry 2002; Boroson 2002). Basing upon the 2-10keV X-ray luminosity, 
Panessa et al. (2006) indicated that the $L/L_{\mathrm{Edd}}$ 
of the Palomar optically selected AGNs ranges from $10^{-5}$ to 0.1. As an additional test, 
the right panel of Figure 4 shows the $M_{\mathrm{BH}}$ vs. $L_{\mathrm{3-20keV}}$ plot. 
$L_{\mathrm{3-20keV}}$ as a function of $M_{\mathrm{BH}}$ is over-plotted as dashed lines for
three different $L/L_{\mathrm{Edd}}$ (i.e., 1, 0.1, and 0.01). As shown in the plot, 
a majority of our objects are located below the line with $L/L_{\mathrm{Edd}}=0.1$ and 
above the line with $L/L_{\mathrm{Edd}}=0.01$.

In summary,  
the hard X-ray selected AGNs listed in our sample are luminous AGNs with a 
wide range of $M_{\mathrm{BH}}$ but a nearly constant accretion activity (i.e., $L/L_{\mathrm{Edd}}$),
which is likely due to the selection effect of the survey that is biased toward 
X-ray luminous objects. In active (or luminous) AGN, the optical/UV ionizing radiation 
is believed to be emitted from a standard geometric thin disk (e.g., Shakura \& Sunyaev 1973). The
low energy photon is comptonized by a hot corona to produce the hard X-ray emission below 10keV (Haardt \& Maraschi 1991). 
A Compton recoil of the soft photon is required to occur on the ionized surface of the accretion disk to produce 
the emission spectrum beyond 10keV. The main observation feature of the reflection is a bump 
peaked at about 30keV 
(e.g., George \& Fabian 1991; Zycki et al. 1994; Ross \& Fabian 2002, 2005).

The bias towards active AGNs could be possibly caused by the fact that 
either intensive Compton reflection
takes place only in AGNs at high state with large $L/L_{\mathrm{Edd}}$ or the surveys are 
biased against the Compton-thick objects. In the first case, the theory of the Compton reflection
predicts that the X-ray emission contributed by 
the reflection depends primarily on the X-ray ionizing parameter, 
and secondarily on the UV radiation produced by the dissipation inside the accretion disk. 
In the second case, our analysis implies a possible connection between the less X-ray absorption
and high $L/L_{\mathrm{Edd}}$ in broad-line AGNs. In fact, we selected the objects with regardless of
their X-ray spectral properties. Figure 1 shows the lack of objects with large 
column density in the sample. 
We believe that the sample completeness would be improved  
by including the ongoing \it Swift\rm/BAT survey with large effective collecting area and 
harder energy bandpass (14-195 keV) in the future studies.     

Figure 4 shows that the hard X-ray luminosity is strongly correlated with the 
$M_{\mathrm{BH}}$ in our sample. In fact, the correlation is naturally expected given the tight 
$L_{\mathrm{3-20keV}}$
vs. $L_{\mathrm{bol}}$ correlation and nearly constant $L/L_{\mathrm{Edd}}$. 
The $L_X$ vs. $M_{\mathrm{BH}}$ correlation provides us a potential estimate of black hole mass for
luminous AGNs within $0.01\leq L/L_{\mathrm{Edd}}\leq 0.1$. The following relationship is obtained 
by us through a least square fitting: 
$\log(M_{\mathrm{BH}}/M_\odot)=(0.82\pm0.09)\log L_{3-20\mathrm{keV}}-(28.23\pm3.78)$. Our results 
conflicts with Panessa et al. (2006) and Pellegrini (2005) who did not find the correlation,
but is in agreement with Kiuchi et al. (2006). Panessa et al. (2006) investigated a sample of 
47 nearby Seyfert galaxies selected from the Palomar spectroscopy (Ho et al. 1997).
The luminosity obtained by 
different X-ray instruments is down to 
$L_{\mathrm{2-10keV}}\sim10^{38} \mathrm{ergs\ s^{-1}}$. 
The studies in Pellegrini (2005) are based on the \it Chandra \rm observations down to a 
luminosity $L_{\mathrm{2-10keV}}\sim10^{38} \mathrm{ergs\ s^{-1}}$.
Kiuchi et al. (2006) used the broad-line AGN sample detected by the \it ASCA \rm Large Sky Survey 
(ALSS) and \it ASCA \rm Medium Sensitivity Survey in the northern sky (AMSSn) with a detection 
limit of a few $\times 10^{-13}\ \mathrm{ergs\ s^{-1}\ cm^{-2}}$ in 2-10keV bandpass. 
As discussed above, it is worthy noting that the correlation most likely does not reflect the 
physics but a selection effect of the surveys.

\subsection{\ion{Fe}{2} ratio vs. disk temperature relation} 

As a key parameter in the EI space, 
the \ion{Fe}{2} ratio (RFe) is defined as the ratio of the optical \ion{Fe}{2} complex to H$\beta$.
Although the total \ion{Fe}{2} emission 
increases by fourfold with reasonable ionization parameter in AGN (Korista et al. 1997), 
traditional photoionization 
models can not explain the strong \ion{Fe}{2} emission in optical and UV bands
(e.g, Netzer \& Wills 1983; Joly 1987; Collin-Souffrin et al. 1988). 
At present, the problem is only slightly alleviated by the major improvements
in the atomic data and by the improved treatment of the line excitation process (Sigut \& Pradhan 2003;
Baldwin et al. 2004). We refer the reads to Collin \& Joly (2000) for a summary of the 
mechanisms that can enhance the \ion{Fe}{2} emission. BG92 put forward a picture in which the RFe is 
determined by the vertical structure of the accretion disk. The vertical structure is 
governed by $L/L_{\mathrm{Edd}}$. A large $L/L_{\mathrm{Edd}}$ leads to a large 
X-ray heated volume that could generate large \ion{Fe}{2} emission. Sulentic et al. (2000b)  
developed a semi-analysis model in which the RFe depends on $L/L_{\mathrm{Edd}}$ as 
$\mathrm{RFe}\propto0.55\log(L/L_{\mathrm{Edd}})$. Although the RFe is found to generally
increase with $L/L_{\mathrm{Edd}}$ in 
optical bright quasars, Netzer et al. (2004) found that a number of high-z 
quasars deviate the trend, i.e., with very small RFe but large $L/L_{\mathrm{Edd}}$ (see also 
in Netzer \& Trakhtenbrot 2007). Using the large database 
provided by the Sloan Digital Sky Survey, Netzer \& Trakhtenbrot (2007) recently suggested that the 
enhanced RFe is mainly caused by increased metal abundance.

In the current sample, the distribution of RFe is shown in the right panel of Figure 5. 
RFe uniformly ranges from $10^{-3}$ to 1. On the contrary, $L/L_{\mathrm{Edd}}$ distributes in a 
quite narrow range as describe above. In fact, no correlation between RFe and $L/L_{\mathrm{Edd}}$
is found in our sample (see the upper panel in Figure 6), which motivates us to suspect that RFe does not 
depend on $L/L_{\mathrm{Edd}}$ only. 
RFe is plotted against the characteristic disk temperature $T_{\mathrm{max}}$ in the 
bottom panel in Figure 6. The temperature scales with $L/L_{\mathrm{Edd}}$ and $M_{\mathrm{BH}}$ as predicted by 
the standard geometric thin disk model (e.g., Shakura \& Sunyaev 1973).
The exact formula of disk temperature depends on various accretion disk models. 
For a rapidly rotating Kerr hole, with a spin parameter $a_*=0.998$ and efficiency of 0.31, we have 
$T_{\mathrm{max}} =10^{5.56} (L/L_{\mathrm{Edd}})^{1/4}M_{\mathrm{BH}}^{-1/4}\ \mathrm{K}$.   
The diagram indicates an obvious correlation between the two parameters.
A spearman rank-order test calculated by survival analysis yields a formal correlation coefficient $r_s=0.414$ 
($P=0.0088$, where $P$ is the probability of null correlation). The correlation is not highly significant 
probably because of the small sample size. The estimated 
temperature spans about one order of magnitude ($\Delta\log T_{\mathrm{max}}\approx1.25$), 
corresponding to a factor of $\sim18$. 
The correlation then suggests a trend of more intensive \ion{Fe}{2} emission for higher disk temperature. 
A marginal dependence of the continuum shape of QSOs on $T_{\mathrm{max}}$ was 
recently identified by Bonning et al. (2007)
who compared the observations of SDSS with the NLTE models of accretion disk. The current result    
means that the strength of the \ion{Fe}{2} emission is likely controlled by the spectral shape of 
the ionizing continuum.  

\subsection{Equivalent width of H$\beta$ vs. $L_X$}
The physical reason of the absence of the Baldwin relationship (Baldwin 1977) for 
low ionization emission lines is still an open question. In fact, a weak inverse Baldwin relationship for H$\beta$
has been demonstrated by recent studies basing upon large AGN samples (e.g., Croom et al. 2002; 
Greene \& Ho 2005). We identify a tight, positive correlation between 
$L_X$ and EW(H$\beta_b$) in our hard X-ray selected AGNs (i.e., an inverse Baldwin relationship).
Figure 7 presents the correlation with correlation coefficient $r=0.611$ ($P=10^{-4}$) estimated by 
the spearman rank-order analysis. 
EW(H$\beta_b$) roughly scales with hard X-ray luminosity as $\mathrm{EW(H\beta_b)}\propto L_X^{0.39}$.
Wilkes et al. (1999) identified a marginal Baldwin effect in H$\beta$ line. Noted that they 
examined only the luminous local quasars with $L_{\mathrm{1-10keV}}>10^{44}\ \mathrm{ergs\ s^{-1}}$.

Although many models are developed to explain the Baldwin effect for high ionization emission 
lines (e.g., \ion{C}{4}, Wandel 1999; Korista et al. 1998; Shields et al. 1995;
Wills et al. 1999; Baskin \& Loar 2004; Bachev et al. 2004), these models can not explain the 
difference between \ion{C}{4} and H$\beta$. Croom et al. (2002) suggested that the inverse 
Baldwin relationship could be explained if the longer wavelength continuum contains emission from 
other components (e.g., thermal dust emission, non-thermal radio emission, starlight). 
We estimate the possible contribution of the unknown sources as follows.
We start from the relationship $\log L_X=39.08+2.56\log \mathrm{EW(H\beta)}$, and re-write 
$\mathrm{EW(H\beta)}=L(H\beta)/(L'+L_c)$, where $L_c$ and $L'$ is the AGN luminosity and 
luminosity of other unknown sources at the H$\beta$ wavelength, respectively. The X-ray luminosity 
could be replaced by $L_c$ given Eq. (2) and the bolometric correction factor of 9.
Replacing $L(H\beta)$ as $L_c$ given Eq.(1) finally yields a relationship 
$\log(1+L'/L_c)=9.48-0.21\log L_c$. Considering the typical case with 
$L_{\mathrm{c}}\sim10^{44}\ \mathrm{erg\ s^{-1}}$, about 40\% of the observed 
continuum at H$\beta$ wavelength is estimated to be contributed by the unknown sources.

\section{Conclusion}

The properties ($L/L_{\mathrm{Edd}}$ and $M_{\mathrm{BH}}$) of accretion onto SMBH are examined in 
a sample of 42 hard X-ray selected (3-60keV) broad-line AGNs in terms of their optical spectra taken by us. 
The energy range is harder than that usually used in the similar previous studies. 
These AGNs are mainly compiled from the RXTE All Sky Survey 
(Sazonov \& Revnivtsev 2004), and are complemented 
by the released \it INTEGRAL \rm AGN sample (Bassani et al. 2006a). 
The statistical analysis allows us to draw the following conclusions: 
\begin{enumerate}
\item We confirm the tight correlation between the hard X-ray and optical emission line luminosities 
(and bolometric luminosity) in our sample, which suggests a close link between the hard X-ray 
emission reflected by the ionized surface of the accretion disk and UV/optical radiation.
Using the hard X-ray luminosity, a strong inverse Baldwin relationship of the H$\beta$ emission line 
is identified in the sample.  
\item The hard X-ray selected broad-line AGNs listed in the sample are found 
to be strongly biased toward luminous AGNs with high $L/L_{\mathrm{Edd}}$ and low column density. 
Since $L/L_{\mathrm{Edd}}$ is constant (mostly between 0.01 and 0.1) in a first order approximation,
the hard X-ray luminosity is strongly correlated with the black hole mass in our sample, which
is most likely due to the selection effect of the surveys.  
 
\item Although the RFe parameter is independent on $L/L_{\mathrm{Edd}}$ in our sample, it is found to be 
correlated with the accretion disk temperature as assessed by $T\propto(L/L_{\mathrm{Edd}})M_{\mathrm{BH}}^{-1}$.
This result implies that the strength of the \ion{Fe}{2} emission is determined by 
the shape of the ionizing spectrum. 
\end{enumerate} 

Finally, it should be mentioned that a new era in AGN hard X-ray study will be opened in next a few years 
due to the launch of new missions with enhanced hard X-ray detection capability in 
not only sensitivity, but also imaging, such as Simbol-X, NeXT and NuSTAR. 
These missions will provide larger, and more complete samples to study the present open issues.








\acknowledgments
We would like to thank the anonymous referee for his/her valuable comments 
that help to improve the paper.  
The authors are grateful to Todd A. Boroson and Richard F. Green for providing 
us the \ion{Fe}{2} template. Special thanks go to the staff at Xinglong 
observatory as a part of National Astronomical Observatories, China Academy of Science 
for their instrumental and observational help. This search is supported by the 
NFS of China under grant 10503005.

\clearpage



\begin{figure}
\plotone{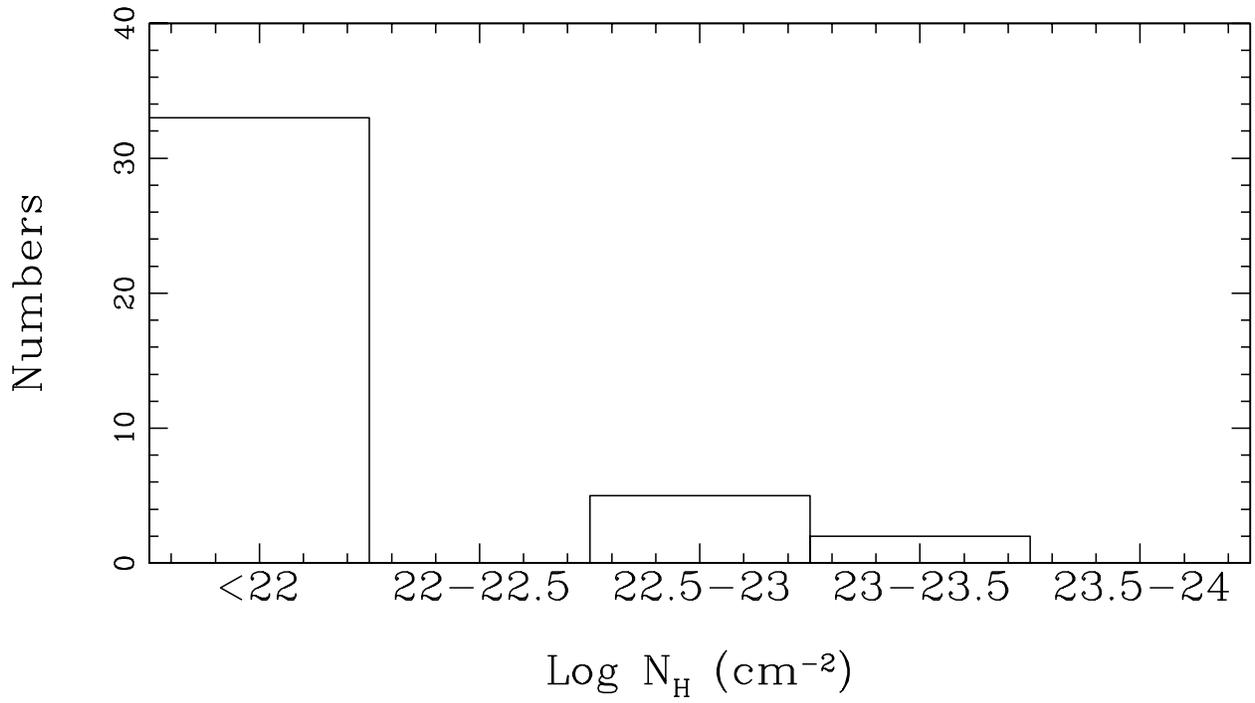}
\caption{Observed X-ray absorption distribution of the 42 hard X-ray selected 
broad-line AGNs listed in the sample. 
}
\end{figure}

\begin{figure}
\plotone{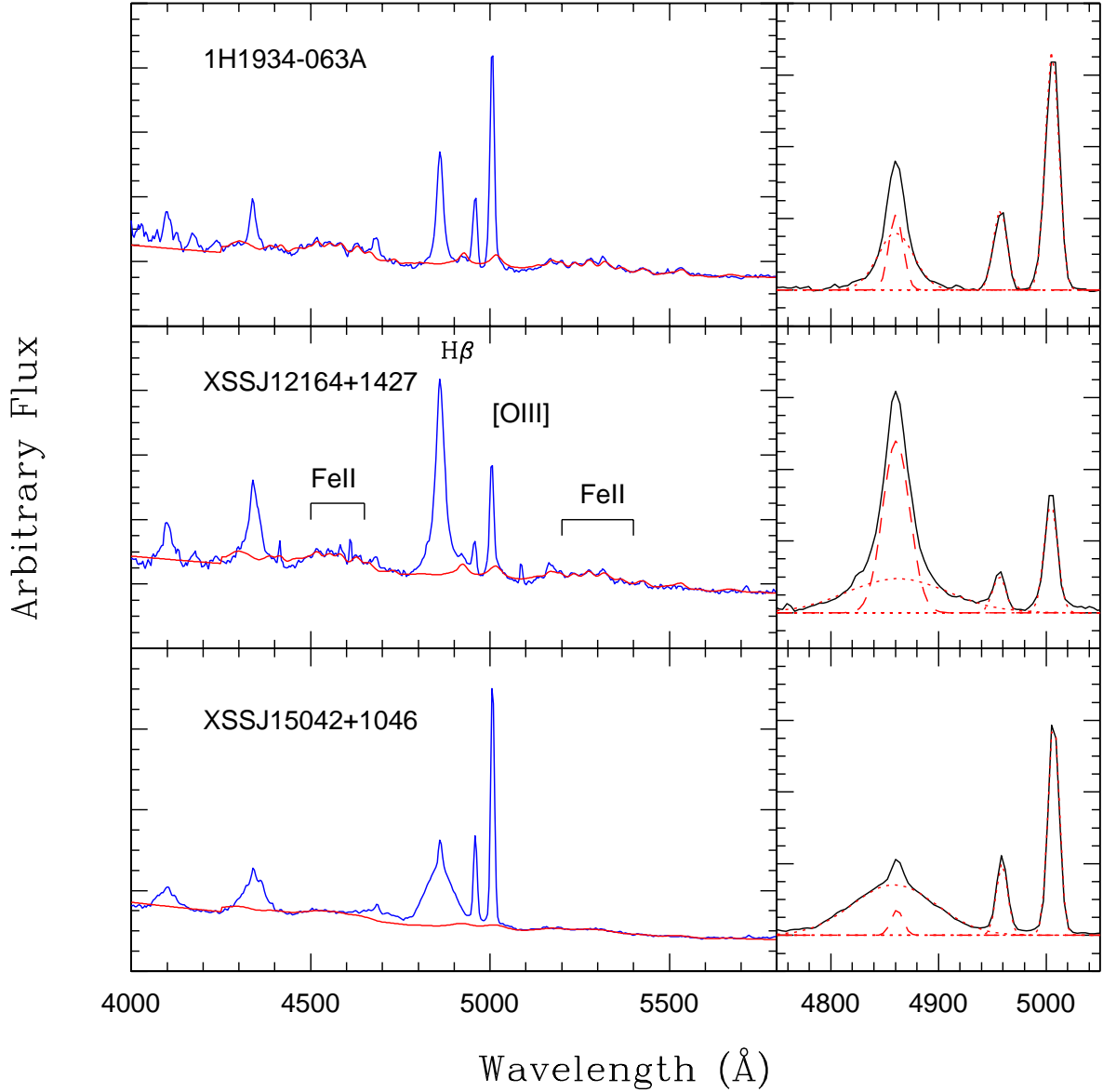}
\caption{\it left column\rm: the modeling of the continuum and \ion{Fe}{2} complex around 
H$\beta$ region for three typical cases. In each panel, the blue curve shows the 
observed spectrum, and the overlaid red one the modeled spectrum (continuum + \ion{Fe}{2} complex).
The strong emission lines, especially the two strong \ion{Fe}{2} complex around the H$\beta$, 
are marked on the figure.
\it right column\rm: the modeling of the emission-line profile
by a set of Gaussian components for the H$\beta$ and [\ion{O}{3}] doublet in 
the three cases, after subtract the modeled continuum and \ion{Fe}{2} complex. In each panel, 
the observed profile is shown by the black curve. The three red short dashed lines show 
the modeled H$\beta$ broad component and [\ion{O}{3}] doublet. The modeled narrow H$\beta$ 
emission is presented by the long dashed line.  
}
\end{figure}

\begin{figure}
\plotone{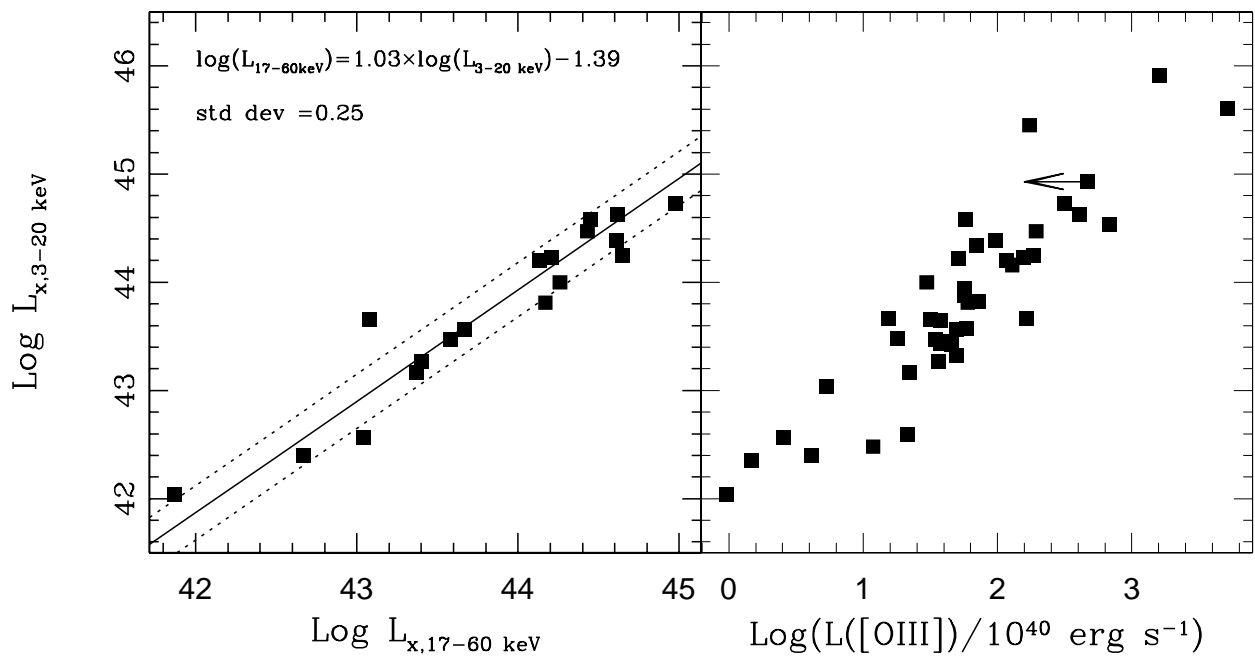}
\caption{ \it left panel\rm: the luminosity in the energy bandpass 3-20 keV plotted  against
the luminosity in 17-60 keV for the 18 objects detected by both XSS and \it INTEGRAL. \rm The solid
line shows the best fit to the data without taking into account of the errors. The $1\sigma$ dispersion
is marked by the two dashed lines. \it right panel\rm: the tight correlation between the 
[\ion{O}{3}] emission line and 3-20 keV luminosity. 
}
\end{figure}

\clearpage
\begin{figure}
\plotone{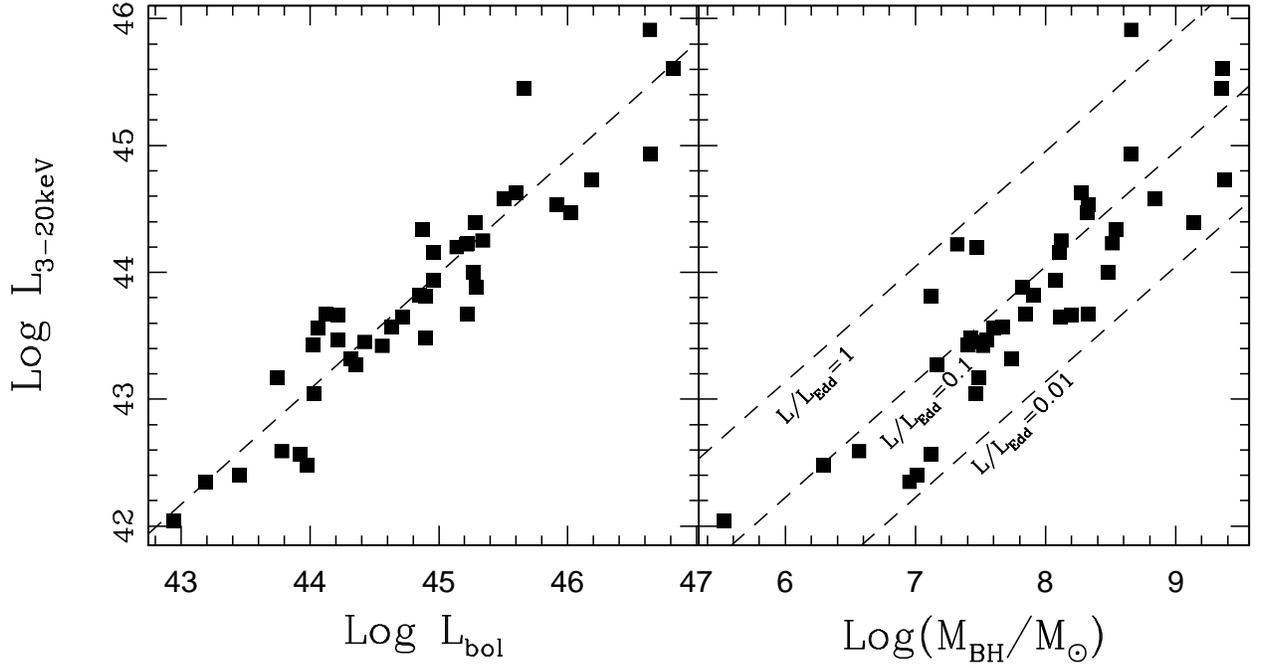}
\caption{\it left panel\rm: $L_{\mathrm{3-20keV}}$ plotted against the bolometric luminosity 
estimated from the H$\beta$ broad component. The solid line shows the best fit to the data: 
$\log L_X = (0.91\pm0.06)\log L_{\mathrm{bol}}+(3.04\pm2.78)$.
\it right panel\rm: Distribution on the plot of $L_X$ vs. black hole mass. The three dashed lines show the 
$L_X$ as a function of the black hole mass for different $L/L_{\mathrm{Edd}}$ =0.01, 0.1, and 1.   
}
\end{figure}

\clearpage
\begin{figure}
\plotone{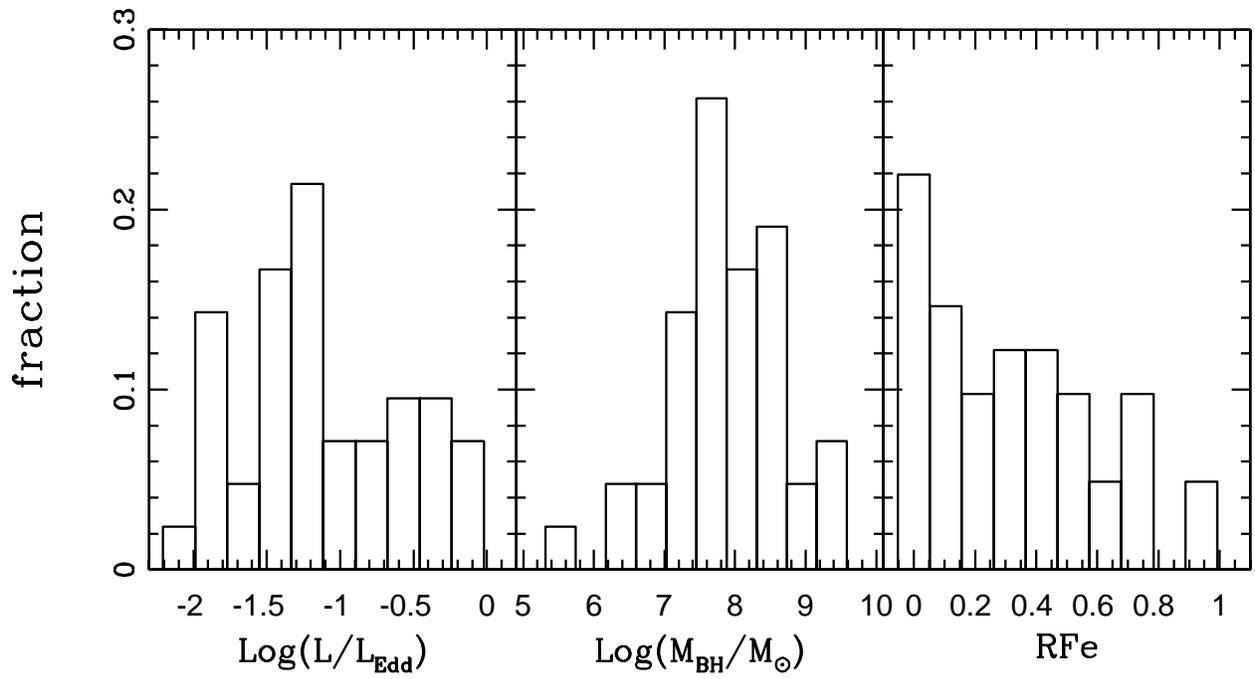}
\caption{Distribution of the Eddington ratio (\it left panel\rm), black hole mass (\it middle panel\rm), 
and parameter RFe (\it right panel\rm).
}
\end{figure}

\clearpage
\begin{figure}
\epsscale{.80}
\plotone{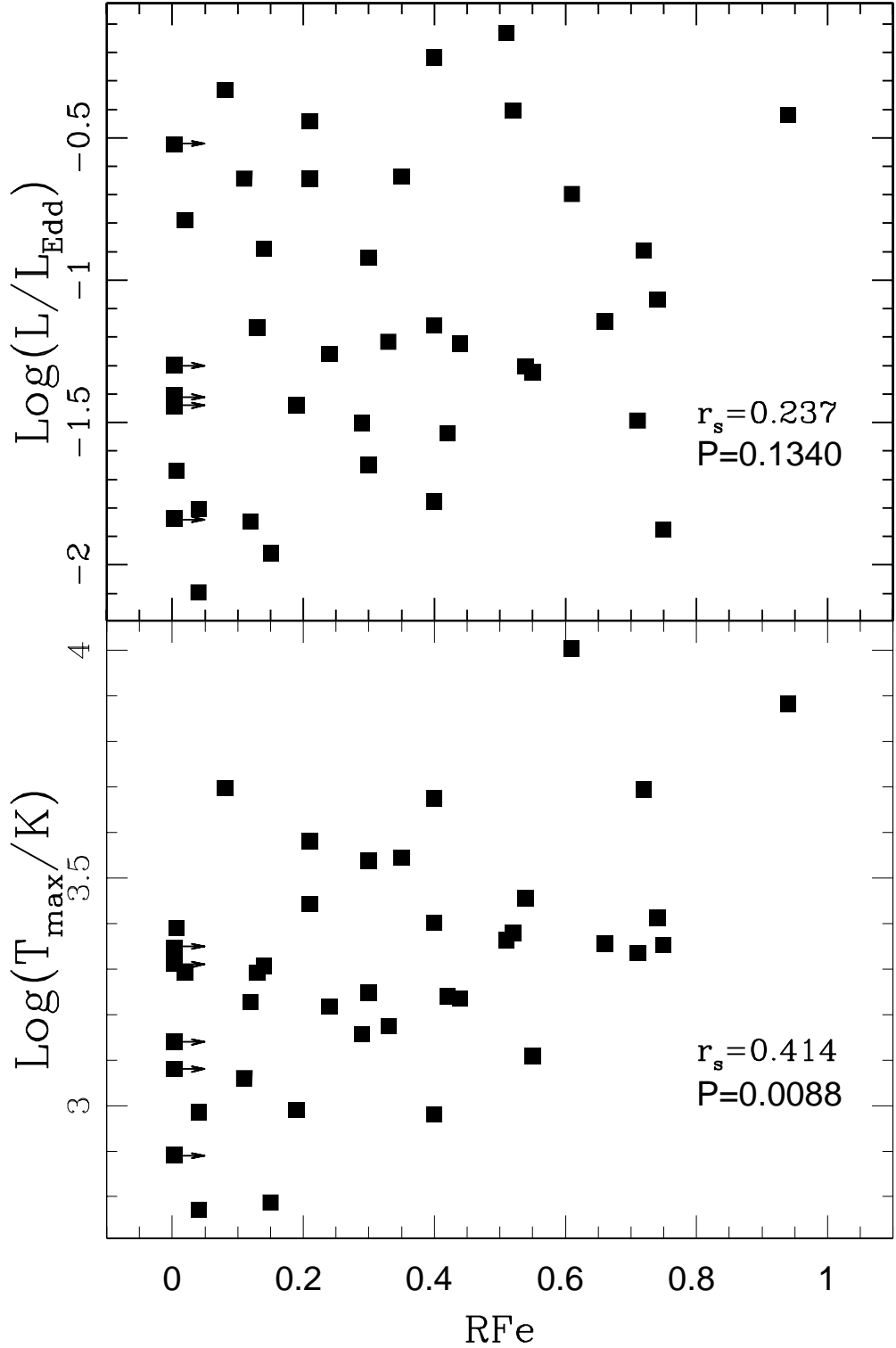}
\caption{\it top panel: \rm RFe plotted against $L/L_{\mathrm{Edd}}$ ($r_s=0.237$, $P=0.134$).  
\it bottom panel: \rm the same as the top panel, but for 
$T_{\mathrm{max}}=10^{5.56} (L/L_{\mathrm{Edd}})^{1/4}{M_{\mathrm{BH}}}^{-1/4}$. 
The diagram shows a relatively strong  correlation between the two parameters ($r_s=0.414$, $P=0.0088$).
In both panels, the points with zero flux of the \ion{Fe}{2} emission are indicated by superposed arrows. 
The correlation coefficients are calculated by survival analysis.  
}
\end{figure}

\clearpage
\begin{figure}
\epsscale{.80}
\plotone{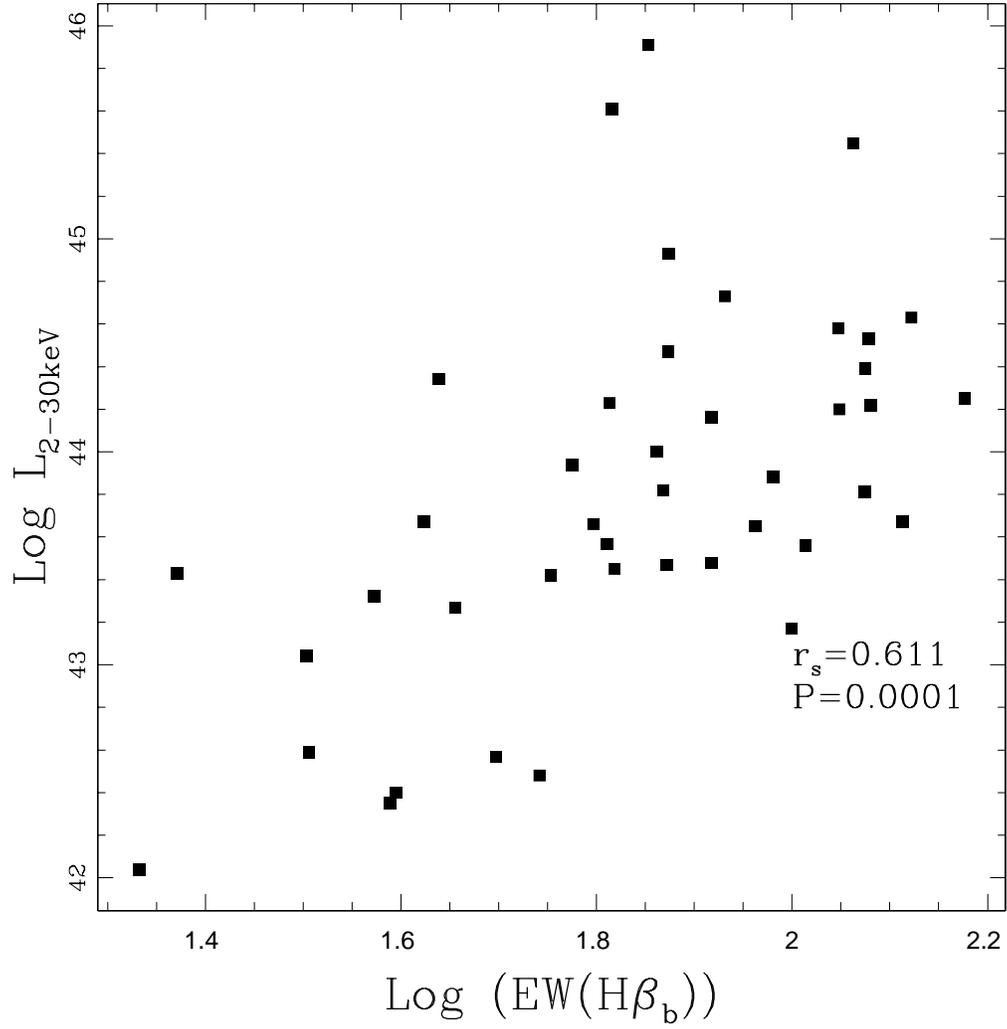}
\caption{Hard X-ray luminosity is plotted as a function of equivalent width of the 
H$\beta$ broad component. A spearman rank-order test yields a correlation coefficient 
$r=0.611$ ($P=10^{-4}$). 
}
\end{figure}








\clearpage



\clearpage

\begin{table}
\scriptsize
\begin{center}
\caption{Log of spectroscopic observation\label{tbl-2}}
\begin{tabular}{ccccccccc}
\tableline\tableline
Name & Other name &  R.A. & Decl. & z & Date & Exposure (s) & $m_v$ & AGN Type \\
(1) & (2) & (3) & (4) & (5) & (6) & (7) & (8) & (9)\\
\tableline
XSS\,J00368+4557 & CGCG\,535-012 & 00 36 21.0 & 45 39 54 & 0.048 & 2400 & 2006 Aug 15 & 15.45 & S1 \\
QSO\,0241+622    & \dotfill  & 02 44 57.7 & 62 28 07 & 0.044 & 3600 & 2006 Dec 17 & 16.1 & S1 \\
3C\,111          & \dotfill & 04 18 21.3 & 38 01 36 & 0.0485 & 3600 & 2007 Nov 03 & 18.0 & S1 \\ 
XSS\,J04331+0520 & 3C\,120 & 04 33 11.1 &  05 21 16 & 0.033 & 1800 & 2007 Feb 16 & 14.2 & BLRG \\    
XSS\,J05103+1640 & IRAS\,05078+1626 & 05 10 45.5 & 16 29 56 & 0.018 & 1800 & 2007 Feb 16 & 15.6 & S1\\
XSS\,J05162-0008 & AKN\,120 & 05 16 11.4 & -00 08 59 & 0.033 & 1200 & 2006 Nov 18 & 14.1 & S1 \\
XSS\,J05552+4617 & MCG\,+8-11-11 & 05 54 53.6 & 46 26 22 & 0.02 & 1800 & 2007 Feb 17 & 15.0 & S1 \\
Mark\,6          & \dotfill & 06 52 12.2 & 74 25 37 & 0.01881 & 3600 & 2007 Feb 16 & 15.0 & S1.5 \\  
XSS\,J07434+4945 & MRK\,79 & 07 42 32.8 & 49 48 35 & 0.022 & 1800 & 2005 Nov 29 & 14.9 & S1 \\
XSS\,J08117+7600 & PG\,0804+761 & 08 10 58.6 & 76 02 42 & 0.1 & 2400 & 2007 Nov 16 & 15.15 & S1\\
XSS\,J09204+1608 & MRK\,704 & 09 18 26.0 & 16 18 19 & 0.029 & 1800 & 2006 Nov 17 & 15.38 & S1 \\
XSS\,J09261+5204 & MRK\,110 & 09 25 12.9 & 52 17 11 & 0.036 & 1800 & 2005 Nov 30 & 15.6  & NLS1 \\
XSS\,J10231+1950 & NGC\,3227 & 10 23 30.6 & 19 51 54 & 0.0038 & 1800 & 2006 Jan 31 & 11.1 &  S1 \\
XSS\,J11067+7234 & NGC\,3516 & 11 06 47.5 & 72 34 07 & 0.008836 & 900 & 2006 Dec 18 & 12.5 & S1 \\
XSS\,J11417+5910 & SBS\,1136+594 & 11 39 08.9 & 59 11 55 & 0.06 & 2400 & 2007 Jan 26 & 15.5 & S1 \\
XSS\,J12032+4424 & NGC\,4051 & 12 03 09.6 & 44 31 53 & 0.0023 & 1200 & 2006 Nov 16 & 10.83 & NLS1 \\
XSS\,J12106+3927 & NGC\,4151 & 12 10 32.6 & 39 24 21 & 0.0033 & 1800 & 2006 Jan 29 & 11.5  & S1 \\
XSS\,J12164+1427 & PG\,1211+143 & 12 14 17.7 & 14 03 13 & 0.081 & 3600 & 2006 Feb 04 & 14.63 & NLS1 \\
NGC\,4253        & \dotfill & 12 18 26.5 & 29 48 46 & 0.01293 & 1200 & 2006 Dec 21 & 13.7 & S1.5 \\
XSS\,J12206+7509 & MRK\,205 & 12 21 44.0 & 75 18 38 & 0.07 & 2400 & 2007 Jan 27 & 14.5 &  S1 \\
XSS\,J12288+0200 & 3C\,273  & 12 29 06.7 & 02 03 08 & 0.15834 & 900 & 2006 Dec 21 & 12.86 & RLQ \\
XSS\,J12408-0516 & NGC\,4593 & 12 39 39.4 & -05 20 39 & 0.009 & 1200 & 2007 Jan 26 & 14.67 & S1 \\
XSS\,J13420-1432 & NPM1G\,-14.0512 & 13 41 12.9 & -14 38 41 & 0.042 & 2400 &  2008 Mar 13 &  15.50 & NLS1 \\ 
XSS\,J13530+6916 & MRK\,279 & 13 53 03.4 & 69 18 30 & 0.031 & 1200 & 2007 Feb 16 & 14.57 & S1 \\
XSS\,J14181+2514 & NGC\,5548 & 14 17 59.5 & 25 08 12 & 0.017 & 1800 & 2008 Mar 03 & 13.3 &  S1 \\
XSS\,J15042+1046 & MRK\,841 & 15 04 01.2 & 10 26 16 & 0.036 & 1200 & 2007 Feb 16 & 14.0 & S1\\
XSS\,J15348+5750 & MRK\,290 & 15 35 52.3 & 57 54 09 & 0.030 & 2400 & 2008 Mar 13 & 15.1 &  S1 \\
XSS\,J17276-1359 & PDS\,456 & 17 28 19.8 & -14 15 56 & 0.184 & 1800 & 2007 Sep 07 & 14.69 & RQQ \\
XSS\,J17413+1851 & 4C\,+18.51 & 17 42 07.0 & 18 27 21 & 0.19 & 2400 & 2007 Sep 07 & 16.43 & RLQ\\
XSS\,J18196+6454 & H\,1821+643 & 18 21 57.3 & 64 20 36 & 0.297 & 1800 & 2006 Aug 16 & 14.1 & RQQ\\
XSS\,J18348+3238 & 3C\,382 & 18 35 02.1 & 32 41 50 & 0.059 & 1800 & 2006 Aug 16 & 15.5 & BLRG\\
XSS\,J18408+7947 & 3C\,390.3 & 18 42 09.0 & 79 46 17 & 0.056 & 1800 & 2006 Aug 15 & 14.37 & BLRG\\
1H\,1934-063A    & \dotfill & 19 37 33.0 & -06 13 05 & 0.01059 & 1800 & 2006 Aug 16 & 14.09 & S1 \\
NGC\,6814        & \dotfill & 19 42 40.6 & -10 19 25 & 0.00521 & 1800 & 2007 Nov 04 & 12.06 & S1.5\\ 
XSS\,J20404+7521 & 4C\,+74.26 & 20 42 37.3 & 75 08 02 & 0.1 & 1800 & 2006 Sep 15 & 15.13 & RLQ\\
XSS\,J20441-1042 & MRK\,509 & 20 44 09.7 & -10 43 25 & 0.034 & 1800 & 2006 Sep 15 & 13.0 & S1\\
XSS\,J21128+8216 & S5\,2116+81 & 21 14 01.2 & 82 04 48 & 0.084 & 2400 & 2006 Aug 15 & 15.7 & BLRG\\
XSS\,J22423+2958 & AKN\,564 & 22 42 39.3 & 29 43 31 & 0.025 & 1800 & 2006 Nov 15 & 14.55 & NLS1 \\
XSS\,J22539-1735 & MR\,2251-178 & 22 54 05.8 & -17 34 55 & 0.064 & 1800 & 2006 Nov 15 & 14.36 & RQQ\\
XSS\,J23033+0858 & NGC\,7469 & 23 03 15.6 & 08 52 26 & 0.016 & 1800 & 2006 Aug 15 & 13.0 &  S1\\
XSS\,J23040-0834 & MRK\,926 & 23 04 43.5 & -08 41 09 & 0.047 & 1800 & 2007 Nov 03 & 14.6 &  S1\\
XSS\,J23073+0447 & PG\,2304+042 & 23 07 02.9 & 04 32 57 & 0.042 & 2400 & 2007 Nov 02 & 15.44 & S1\\  
 
\tableline
\end{tabular}
\end{center}
\end{table}



\begin{deluxetable}{ccccccccccc}
\tabletypesize{\scriptsize}
\rotate
\tablecaption{List of properties of the hard X-ray selected AGNs}
\tablewidth{0pt}
\tablehead{ 
\colhead{Name} & \colhead{$\mathrm{EW(H\beta_n)}$} & \colhead{$\mathrm{EW(H\beta_b)}$}  
& \colhead{RFe} & \colhead{$\log[L(H\beta_b)]$} & \colhead{FWHM(H$\beta_b$)} & \colhead{$\log[L([\mathrm{OIII}])]$} & 
\colhead{$\log[L_{2-30\mathrm{keV}}]$} & \colhead{$\log[L_{17-60\mathrm{keV}}]$} & \colhead{
$\log(M_{\mathrm{BH}}/M_\odot)$} & \colhead{$L/L_{\mathrm{Edd}}$}\\
\colhead{} & \colhead{\AA} & \colhead{\AA} & \colhead{} & \colhead{$\mathrm{ergs\ s^{-1}}$} 
& \colhead{$\mathrm{km\ s^{-1}}$} & \colhead{$\mathrm{ergs\ s^{-1}}$} & \colhead{$\mathrm{ergs\ s^{-1}}$} 
& \colhead{$\mathrm{ergs\ s^{-1}}$} & \colhead{} & \colhead{}\\ 
\colhead{(1)} & \colhead{(2)} & \colhead{(3)} & \colhead{(4)} & \colhead{(5)} & \colhead{(6)} &
\colhead{(7)} & \colhead{(8)} & \colhead{(9)} & \colhead{(10)} & \colhead{(11)}  \\
}
\startdata
XSSJ\,00368+4557   &   \dotfill &  82.6  &  0.35  & 42.1 &  2555  &  41.3  & 43.48 & \dotfill &    7.42 & 0.23 \\
QSO\,0241+622      &   \dotfill &  43.6  &  0.26  & 42.1 &  9401  &  41.8  & 44.34 & 44.28    &    8.54 & 0.02 \\
3C\,111            &   \dotfill & 119.9  &  0.00  & 43.3 &  3445  &  42.8  & 44.53 & 44.59    &    8.33 & 0.30 \\
XSS\,J04331+0520   &   10.5     & 82.7   &  0.24  & 42.2 &  5367  &  42.1  & 44.16 & \dotfill &    8.11 & 0.06 \\  
XSS\,J05103+1640   &    9.0     & 74.5   &  $>0.$ & 41.3 &  4855  &  41.5  & 43.47 & 43.58    &    7.55 & 0.04\\
XSS\,J05162-0008   &    2.3     & 72.7   &  0.55  & 42.5 &  6595  &  41.5  & 44.   & 44.26    &    8.48 & 0.05\\
XSS\,J05552+4617   &   53.5     & 103.2  &  0.30  & 41.2 &  5761  &  41.7  & 43.56 & 43.67    &    7.60 & 0.02\\
Mark\,6            &    6.0     & 37.4   &  0.42  & 41.4 &  5627  &  41.7  & 43.32 & 43.41    &    7.74 & 0.03\\
XSS\,J07434+4945   &    8.6     & 65.8   &  0.40  & 41.6 &  3815  &  41.7  & 43.45 & \dotfill &    7.47 & 0.07\\
XSS\,J08117+7600   &  \dotfill  & 74.7   &  0.52  & 43.4 &  3146  &  42.3  & 44.47 & 44.43    &    8.32 & 0.39\\
XSS\,J09204+1608   &    6.9     & 91.8   &  0.29  & 41.9 &  6421  &  41.6  & 43.65 & \dotfill &    8.11 & 0.03\\
XSS\,J09261+5204   &  \dotfill  & 118.6  &  0.08  & 42.1 &  1798  &  41.8  & 43.81 & 44.17    &    7.12 & 0.47\\
XSS\,J10231+1950   &    5.1     & 39.3   &  0.007 & 40.5 &  4556  &  40.6  & 42.4  & 42.67    &    7.01 & 0.02\\
XSS\,J11067+7234   &  \dotfill  & 31.9   &  \dotfill & 41.1 & 5020 & 40.7  & 43.04 & \dotfill &    7.46 & 0.03\\ 
XSS\,J11417+5910   &    7.3     & 73.8   &  0.13  & 42.0 &  4608  &  41.9  & 43.82 & \dotfill &    7.91 & 0.07\\
XSS\,J12032+4424   &  \dotfill  & 21.5   &  0.61  & 39.9 &  1202  &  40.0  & 42.04 & 41.87    &    5.53 & 0.20\\
XSS\,J12106+3927   &   37.2     & 100.0  &  0.12  & 40.8 &  6354  &  41.3  & 43.17 & 43.37    &    7.49 & 0.01 \\
XSS\,J12164+1427   &  \dotfill  & 120.3  &  0.40  & 42.5 &  1804  &  41.7  & 44.22 & \dotfill &    7.32 & 0.60\\
NGC\,4253          &  11.3      & 32.1   &  0.72  & 40.8 &  2153  &  41.3  & 42.59 & 42.7     &    6.57 & 0.13\\
XSS\,J12206+7509   &  \dotfill  & 95.6   &  0.21  & 42.5 &  3037  &  41.8  & 43.88 & \dotfill &    7.83 & 0.23\\
XSS\,J12288+0200   &  \dotfill  & 71.3   &  0.51  & 44.1 &  2970  &  43.2  & 45.91 & 45.92    &    8.66 & 0.74\\
XSS\,J12408-0516   &  \dotfill  & 49.8   &  0.54  & 41.0 &  3654  &  40.4  & 42.57 & 43.04    &    7.12 & 0.05\\
XSS\,J13420-1432   &   15.3     & 64.6   &  0.66  & 41.8 &  4112  &  41.8  & 43.57 & \dotfill &   7.67 & 0.07\\
XSS\,J13530+6916   &    5.7     & 59.6   &  0.44  & 42.2 &  5147  &  41.8  & 43.94 & \dotfill &    8.07 & 0.06\\
XSS\,J14181+2514   &    7.8     & 62.7   &  0.04  & 41.3 & 10312  &  41.5  & 43.66 & 43.08    &    8.20 & 0.008\\
XSS\,J15042+1046   &    9.4     & 129.6  &  0.33  & 42.5 &  5709  &  42.2  & 43.67 & \dotfill &    8.33 & 0.061\\
XSS\,J15348+5750   &    4.2     & 23.5   &  0.71  & 41.1 &  4740  &  41.6  & 43.43 & \dotfill &    7.41 & 0.03\\
XSS\,J17276-1359   &  \dotfill  & 74.6   &  3.58  & 44.1 &  2957  &  $<42.7$ & 44.93 & \dotfill &  8.66 & 0.75\\  
XSS\,J17413+1851   &    1.8     & 115.5  &  0.04  & 43.0 & 13506  &  42.2  & 45.45 & \dotfill &    9.35 & 0.02\\
XSS\,J18196+6454   &    2.9     & 65.4   &  0.18  & 44.3 &  5797  &  43.7  & 45.61 & \dotfill &    9.36 & 0.23\\
XSS\,J18348+3238   &  \dotfill  & 111.4  &  0.19  & 42.8 &  8340  &  41.8  & 44.58 & 44.45    &    8.84 & 0.04\\
XSS\,J18408+7947   &    3.2     & 118.8  &  0.15  & 42.5 & 13799  &  42.0  & 44.39 & 44.61    &    9.13 & 0.01\\
1H\,1934-063A      &  \dotfill  & 55.2   &  0.94  & 41.1 &  1354  &  41.0  & 42.48 & 42.59    &    6.29 & 0.38\\
NGC\,6814          &  \dotfill  & 38.8   &  0.75  & 40.2 &  5182  &  40.2  & 42.35 & 42.47    &    6.95 & 0.01\\
XSS\,J20404+7521   &  \dotfill  & 85.4   &  $>0.$ & 43.6 &  9418  &  42.5  & 44.73 & 44.98    &    9.37 & 0.05\\
XSS\,J20441-1042   &   10.7     & 111.8  &  0.21  & 42.4 &  2261  &  42.1  & 44.2  & 44.13    &    7.47 & 0.36\\
XSS\,J21128+8216   &    6.3     & 150.0  &  0.14  & 42.6 &  4123  &  42.3  & 44.25 & 44.65    &    8.12 & 0.13\\
XSS\,J22423+2958   &  \dotfill  & 56.6   &  0.74  & 41.7 &  3649  &  41.7  & 43.42 & \dotfill &    7.52 & 0.09\\
XSS\,J22539-1735   &    5.7     & 132.5  &  0.02  & 42.9 &  4093  &  42.6  & 44.63 & 44.62    &    8.28 & 0.16\\
XSS\,J23033+0858   &   13.7     & 45.3   &  0.30  & 41.5 &  2819  &  41.6  & 43.27 & 43.40    &    7.17 & 0.12\\
XSS\,J23040-0834   &    3.4     & 65.1   &  $>0.$ & 42.5 &  7083  &  42.2  & 44.23 & 44.21    &    8.51 & 0.04\\
XSS\,J23073+0447   &    2.7     & 42.0   &  $>0.$ & 41.2 &  7322  &  41.2  & 43.67 & \dotfill &    7.85 & 0.01\\
\enddata
\end{deluxetable}




\end{document}